\documentclass[twoside]{article}
\usepackage[papersize={5.5in,8.27654in},total={4.25in,7.25in},%
  includehead,nofoot,centering]{geometry}
\usepackage{amsmath,txfonts}
    \DeclareSymbolFont{cmmath}{OML}{mlmm}{m}{it}
    \DeclareMathSymbol{v}{\mathord}{cmmath}{`v}
    \DeclareMathSymbol{\nu}{\mathord}{cmmath}{23}
\usepackage[T1]{fontenc}
\usepackage[colorlinks=true,allcolors=blue,bookmarks=false,
  pdfauthor={Gavin R. Putland},
  pdftitle={Exact formulation of Huygens' principle in terms of generalized spatiotemporal-dipole secondary sources},
  pdfkeywords={wave optics, diffraction, Kirchhoff's integral theorem, Huygens' principle, Fermat's principle, obliquity factor, inclination factor, Huygens-Kirchhoff, Fresnel-Kirchhoff, Huygens-Fresnel},
  pdfsubject={Optics, Physical Optics, Diffraction}]{hyperref}
\usepackage{tikz}
  \tikzset{>=latex} 
\usepackage[small]{caption}  

\usepackage[final,kerning,spacing]{microtype}
  \SetExtraKerning{encoding={*}}{1={-35,-45}, / ={50,50}, ( ={,-30}, ) ={-30,},
    \textemdash={50,50}, \textendash={30,30}, \textquotedblleft={-30,}}
  \SetExtraSpacing{encoding={*}}{A={-80,,}, f={80,,}, r={-80,,}, v={-80,,},
    w ={-70,,}, y={-60,,}, ' = {-200,,}, \textquotedblright={-80,,}}

\title{\LARGE Exact formulation of Huygens' principle\\\large in terms of\Large\\generalized spatiotemporal-dipole secondary sources}
\author{Gavin R.{\small~}Putland~\!\thanks{\,\small Royal Melbourne Institute of Technology, Australia.~ Gmail name: grputland.\\Copyright license: \href{https://creativecommons.org/licenses/by/4.0/legalcode}{Creative Commons Attribution 4.0 International}.}}
\date{\vspace{-1ex}\normalsize 8 October 2025} 

\makeatletter
\def\ps@plain{
  \let\@oddfoot=\@empty
  \def\@oddhead{\hfill\normalsize\sf\thepage}
  }
\def\ps@headings{
  \def\sectionmark##1{\markright{\S\thesection.~ ##1}}
  \let\@evenfoot=\@empty  \let\@oddfoot=\@empty
  \def\@evenhead{\footnotesize\sf\underline{\makebox[\textwidth]{\normalsize\sf\thepage\hfill\footnotesize\sf Putland,\itshape\, Exact formulation of Huygens' principle~\!...}}}
  \def\@oddhead{\footnotesize\sf\underline{\makebox[\textwidth]{\rightmark\hfill\normalsize\sf\thepage}}}
  }
\makeatother

\newcommand{\thefigurename}{Fig.}
\makeatletter
  \def\fnum@figure{\textbf{\thefigurename\,\thefigure}}
\makeatother

\begin{document}

\sloppy

\maketitle

{\small

\subsubsection*{Abstract}

\indent\indent A ``spatiotemporal dipole'' wave source, as defined by
D.~\!A.B.\,Miller (1991), differs from an ordinary (``spatial'')
dipole source in that the inverted monopole is delayed relative to the
uninverted monopole, the delay being equal to the propagation time
from one monopole to the other. \mbox{A ``generalized''}
spatiotemporal dipole (GSTD), as defined here, is generalized in two
ways: first, the delay may be smaller in absolute value (but not
larger) than the propagation time, so~that the radiated waves cancel
at a certain angle from the axis of the dipole; second, one monopole
may be attenuated relative to the other, so~that the cancellation is
exact at a finite distance\textemdash on a circle coaxial with the
dipole.

I show that the Kirchhoff integral theorem, for a single monopole
primary source, gives the same wave function as a certain distribution
of GSTD secondary sources on the surface of integration. In the GSTDs,
the ``generalized'' delay allows the surface of integration to be
general (not necessarily a primary wavefront), whereas the attenuation
allows an exact match of the wave function even in the near field of
the primary source. At each point on the surface of integration, the
circle of cancellation of the GSTD secondary source passes through the
primary source, which therefore receives no backward secondary waves,
while the direction of specular reflection of the primary wave passes
through the same circle, giving a geometrical-optical explanation of
the suppression of backward secondary waves at any field point.

}

\clearpage

\pagestyle{headings}

\section{Background}

Let $R$ be the region inside a closed surface~$S$, and let $R'$ be the
region outside~$S$.  Let the ``wave function'' $\psi(P,t)$, at a
general ``field point''~$P$~\! and a general time~$t$, be a solution
of the wave equation {\large(}${\ddot{\psi}=c^2\;\!\nabla^2\psi}$~\!
for constant~$c${\large)} inside~$R$, due to sources in~$R'$
(satisfaction of the wave equation in~$R$ means there are no sources
in~$R$). Let $n$ be the \emph{normal coordinate} measured from the
surface~$S$ into~$R$, and let $s$ be the distance from $P$~\! to the
general element of the surface~$S$, with area~$dS$ (so~that $s$ can be
considered a coordinate of the surface element, measured
from~$P$). Then, according to the \textbf{Kirchhoff integral theorem},
\begin{equation}
  \frac{1}{4\pi}\! \iint\limits_{\!S} \!\Bigg\{
    [\psi]~\!\frac{\partial}{\partial n}\bigg(\frac{\,1\,}{s}\bigg)
    - \frac{1}{cs}\big[\dot{\psi}\big] ~\!\frac{\partial s}{\partial n}
    - \frac{\,1\,}{s}\bigg[\frac{\partial\psi}{\partial n}\bigg]
  \Bigg\} \,dS
  =\,
  \begin{cases}
    \psi(P,t) \!\! & \text{for } P \text{ in } R\\[1ex]
    0              & \text{for } P \text{ in } R',
  \end{cases}\label{e-kit}
\end{equation}
where derivatives w.r.t.~$n$ are taken at the surface element
(i.e.\,at ${n\!=\!0}$), and square brackets indicate that the enclosed
function is evaluated at the surface element, at time\, ${t-s/c\;\!}$
(that~is, \emph{delayed} by the propagation time
to~$P$).\footnote{\,\small Born \& Wolf~\cite{born-wolf-02} at
pp.\,420--21, especially eq.\,(13).\, \textit{Cf}.~Baker \&
Copson~\cite[p.~\!37]{baker-copson-39} and Miller
\cite[eq.\,2]{miller-91}, who use $r$ instead of\,$s$ (among other
notational differences). Later, Baker \& Copson change the
sign \cite[p.~\!40, last~eq.]{baker-copson-39} because they use a new
normal coordinate~$\nu$, measured in the opposite direction.}

The theorem can be extended to an infinite region~$R$ by adding
another sheet to the bounding surface~$S$ in such a way that the
additional sheet, at least in the limiting case, makes no contribution
to the surface integral. One way to satisfy the latter requirement is
to suppose that the additional sheet is so far away that the
disturbance has not reached it yet! If~that is not permissible (e.g.,
due to strict sinusoidal time-dependence), we~can consider how the
wave function decays with
distance~\cite[pp.\,37--8]{baker-copson-39}. By such methods we can
apply~\eqref{e-kit} not only to the region inside a closed surface,
but also (e.g.) to the region outside a closed surface, or the region
on one side of an infinite open surface.\footnote{\,\small The
derivation in~\cite{putland-22} is indifferent to such distinctions.}

\emph{If} the integral in~\eqref{e-kit} can be interpreted as the wave
function at~$P$~\! due to a distribution of ``secondary'' sources
on~$S$, the theorem will show that~\! (i)~the wave function in~$R$
is \emph{as~if} the ``primary'' sources in~$R'$ were \emph{replaced}
by the said distribution of ``secondary'' sources on $S$, but~\!
(ii)~the wave function in~$R'$ due to the same distribution of
``secondary'' sources is null\textemdash in other words, the secondary
sources collectively give \emph{no backward secondary waves}.  Thus we
will have successfully mathematized \textbf{Huygens' principle}.

One such interpretation of the integral is well known\textemdash and,
for present purposes, worth deriving in detail. If~the
time-dependences indicated by the square brackets in~\eqref{e-kit} are
made explicit (while the spatial variations are left \emph{im}plicit),
the integrand becomes
\begin{equation}\label{e-kand1}
  \psi\big(t-s/c\big)\,\frac{\partial}{\partial n}\bigg(\frac{\,1\,}{s}\bigg)
  \,-\, \frac{1}{cs}~\!\psi'\big(t-s/c\big)~\!\frac{\partial s}{\partial n}
  \,-\, \frac{\,1\,}{s}~\!\tfrac{\partial\psi}{\partial n}\big(t-s/c\big) \,.
\end{equation}
In the third term, as implied in~\eqref{e-kit},\,
$\frac{\partial\psi}{\partial n}$ does not allow for the variation
in~$s$ with~$n$; rather, ${\frac{\partial\psi}{\partial
n}\big(t\big)}$ is evaluated at~${n\!=\!0}$, ignoring~$s$, and then
delayed by~$s$/$c$. But in the second and first terms, the operator
$\frac{\partial}{\partial n}$ is applied directly to $s$ and its
reciprocal, and therefore obviously \emph{does} count the variation
in~$s$ with~$n$. Using the chain rule twice, the second term
in~\eqref{e-kand1} with its leading sign can be written
\begin{align}
  \frac{\,1\,}{s}\,\psi'\big(t-s/c\big)\bigg(\frac{-1}{c}\bigg)
    \,\frac{\partial s}{\partial n} \,
&=\, \frac{\,1\,}{s}\,\frac{\partial}{\partial s}\Big(\psi\big(t-s/c\big)\Big)
    \,\frac{\partial s}{\partial n} \nonumber\\[.5ex]
&=\, \frac{\,1\,}{s}\,\frac{\partial}{\partial n}\Big(\psi\big(t-s/c\big)\Big)\,.
\end{align}
Substituting this back into~\eqref{e-kand1}, and recognizing the first
two terms as the derivative of a product, we~find that the integrand
in~\eqref{e-kit}~is
\begin{equation}\label{e-kand2}
  \frac{\partial}{\partial n}\bigg(\frac{\,1\,}{s}\,\psi\big(t-s/c\big)\bigg)
  \,-\, \frac{\,1\,}{s}~\!\tfrac{\partial\psi}{\partial n}\big(t-s/c\big) \,.
\end{equation}
In \eqref{e-kand2}, the second term with its leading sign is
recognizable as the wave function due to a monopole source with
strength $-\frac{\partial\psi}{\partial n}$ [per unit area, divided by
$4\pi$ in~\eqref{e-kit}],\footnote{\,\small For better or worse,
I~follow Baker \& Copson~\cite[p.\,42]{baker-copson-39}, Born \&
Wolf~\cite[p.\,421]{born-wolf-02}, and
Larmor~\cite[p.~\!244]{larmor-1929} in defining \textit{strength} so
that the wave function at distance~$s$ from a monopole source with
strength~$f(t)$\, is $\frac{f(t-s/c)}{s}$, omitting the factor~$4\pi$
from the denominator. Miller~\cite[p.~\!1371]{miller-91} includes this
factor.} whereas the first term can be conveniently written
\begin{equation}\label{e-ch}
  h\,\frac{\partial}{\partial n}
  \bigg(\frac{\,1\,}{s}~\!\frac{~\!\psi\big(t-s/c\big)}{h}\bigg)
\end{equation}
for infinitesimal $h$. Here the expression in the big parentheses is
the wave function due to a monopole source with strength
$\frac{\psi(t)}{h}$, so~that the complete expression~\eqref{e-ch} is
the \emph{change} in that wave function due to shifting that source
from (say) ${n\!=\!-h}$~\! to ${n\!=\!0}$. Thus
expression~\eqref{e-ch} is the wave function due to a composite source
consisting of an ``uninverted'' monopole with strength
$\frac{\psi(t)}{h}$~\! at ${n\!=\!0}$, and an ``inverted'' monopole
with strength $\frac{-\psi(t)}{h}$~\! at ${n\!=\!-h}$, for
infinitesimal~$h\;\!$; we~call this combination a \textit{dipole}
(or~\textit{doublet}) whose strength (or~\textit{moment}) is
$\psi(t)$, in the $n$~direction\textemdash the \emph{normal}
direction.

So the complete integrand in~\eqref{e-kit} describes a monopole
secondary source with strength $-\frac{\partial\psi}{\partial n}$ and
a normal dipole secondary source with strength $\psi(t)$, per unit
area. This interpretation is long
established~\cite[pp.\,42--3]{baker-copson-39}. But, as noted by
David\,A.B.\,Miller~\cite[p.~\!1370]{miller-91}, it~is problematic in
that each element of the surface~$S$ corresponds to \emph{two}
secondary sources instead of one. For the purpose of quantifying
Huygens' principle, it~would be preferable to express the combination
as a \emph{single, directional} secondary source, whose directionality
relates to the suppression of backward secondary waves.

\section{Generalized spatiotemporal dipoles (GSTDs)}

A single source matching the integrand in~\eqref{e-kit}, can be found
by a na\"{i}ve method: generalize the dipole in~\eqref{e-ch},
introducing undetermined parameters, and then adjust the parameters
so~that the wave function agrees with~\eqref{e-kand2}. The obvious way
to generalize the dipole (I~thought) is to introduce an adjustable
delay between the monopoles, not exceeding the propagation time
between them, so~that the radiated waves interfere destructively at an
adjustable angle from the $n$~direction (the axis of the dipole). But
this turns out to give too few parameters to match the coefficients,
even in the simple case of a single monopole primary source. We~can
inject a second unknown by attenuating one monopole by an adjustable
fraction, so~that the radiated waves cancel exactly at an adjustable
distance in the direction of destructive interference.

So let us modify the spatial dipole by delaying the strength function
of the inverted monopole by~$\tau_h$, and reducing its magnitude by
the fraction $\alpha_h$ (no~reduction if~$\alpha_h\!=\!0$, complete
nullification if~$\alpha_h\!=\!1$). Then, compared with the uninverted
monopole, the inverted monopole is recessed by the distance $h$,
delayed by the time $\tau_h$, and attenuated by the fraction
$\alpha_h$. Recall that the wave function at~$P$~\! due to
the \emph{un}inverted monopole, in the big parentheses
in~\eqref{e-ch}, is
\begin{equation}\label{e-uninv}
  \frac{\psi\big(t-s/c\big)}{hs} \,.
\end{equation}
So the wave function due to the modified dipole is the total change
in~\eqref{e-uninv} due to $n$ increasing by $h$,\, and $t$ increasing
by $\tau_h\;\!$,\footnote{\,\small If we introduce a delay $u$, the
numerator of~\eqref{e-uninv} becomes ${\psi\big(t-u-s/c\big).}$ In the
change from the inverted monopole to the uninverted monopole,
$u$~falls from $\tau_h$ to~$0$, which has the same effect on the
function as if $t$ \emph{increases} by $\tau_h$.} and the magnitude
increasing by $\alpha_h$ times its final value. Since $h$ and~$\tau_h$
are infinitesimal, that total change is
\begin{equation}\label{e-tot-ch}
  h~\!\frac{\partial}{\partial n}\bigg(\frac{\psi\big(t-s/c\big)}{hs}\bigg)\,+\,
  \tau_h \,\frac{\partial}{\partial t}\bigg(\frac{\psi\big(t-s/c\big)}{hs}\bigg)
  \,+\, \alpha_h~\!\frac{\psi\big(t-s/c\big)}{hs} ~,
\end{equation}
i.e.
\begin{equation}
  \frac{\partial}{\partial n}\bigg(\frac{\psi\big(t-s/c\big)}{s}\bigg) \,+\,
  \frac{\,\tau_h\,}{hs} ~\!\dot{\psi}\big(t-s/c\big) \,+\,
  \frac{\,\alpha_h\,}{hs} ~\!\psi\big(t-s/c\big) \,,
\end{equation}
which will agree identically with~\eqref{e-kand2} if and only if,
on~$S$,
\begin{equation}\label{e-mod-match}
  \frac{\,\tau_h\,}{h}~\!\dot{\psi} \,+\, \frac{\,\alpha_h\,}{h}~\!\psi
  \,=\, -\frac{\partial\psi}{\partial n} \,.
\end{equation}
This is the sufficient and necessary condition for the modified
dipoles, and the original dipoles and monopoles, to give identical
secondary waves.

As condition~\eqref{e-mod-match} is a simple linear dependence between
a wave function~$\psi$, its time-derivative, and one of its
directional derivatives, we~should not expect to be able to satisfy it
for a general wave function, but \emph{should} expect that we can
satisfy it for a particular direction of propagation. So let us take
the special case of a \textbf{single monopole primary source}, with
strength $f(t)$, located at point~$O$~\! in~$R'$. If the
coordinate~$r$ is the distance from this source, then the primary wave
function~is
\begin{equation}\label{e-psimonop}
  \psi = \frac{\,1\,}{r}\;\! f\big(t-r/c\big) \,.
\end{equation}
By comparing the partial derivatives of this wave function
w.r.t.~$r$ and $t$, we readily obtain the relation
\begin{equation}\label{e-diffs}
  \frac{\partial\psi}{\partial r}
  \,=\, {-}\frac{\,\dot{\psi}\,}{c} - \frac{\,\psi\,}{r} \,.
\end{equation}

Now we can apply condition~\eqref{e-mod-match}. Considering $r$ as
a function of $n$ for each element of~$S$, we~can use the chain rule
on the right of~\eqref{e-mod-match}, obtaining
\begin{equation}\label{e-chain}
  \frac{\,\tau_h\,}{h}~\!\dot{\psi} \,+\, \frac{\,\alpha_h\,}{h}~\!\psi
  \,=\, -\frac{\partial\psi}{\partial r}~\!\frac{\partial r}{\partial n} \,.
\end{equation}
But, by the geometry,
\begin{equation}\label{e-drdn}
  \frac{\partial r}{\partial n} = \cos(n,r) \,,
\end{equation}
in which the right-hand side is the cosine of the angle between the
positive directions of $n$ and~$r$. Substituting \eqref{e-diffs}
and~\eqref{e-drdn} into~\eqref{e-chain} gives
\begin{equation}\label{e-tauh-alphah}
  \frac{\,\tau_h\,}{h}~\!\dot{\psi} \,+\, \frac{\,\alpha_h\,}{h}~\!\psi
  \,=\, \bigg(\frac{\,\dot{\psi}\,}{c} + \frac{\,\psi\,}{r}\bigg)\,\cos(n,r)\,.
\end{equation}
To satisfy this for all $\psi$ of the form~\eqref{e-psimonop}, we
equate the coefficients of $\dot{\psi}$, and equate the coefficients
of $\psi$, obtaining respectively
\begin{align}
  \tau_h   &= \tfrac{\,h\,}{c}\cos(n,r) \,, \label{e-tauh} \\[.5ex]
  \alpha_h &= \tfrac{\,h\,}{r}\cos(n,r) \,, \label{e-alphah}
\end{align}
so that the parameters of the ``modified'' dipole are uniquely
determined. Substituting \eqref{e-tauh} and~\eqref{e-alphah}
into~\eqref{e-tot-ch} and collecting the operators, we~find that the
integrand in~\eqref{e-kit} becomes
\begin{equation}\label{e-gstd}
  \Bigg\{\frac{\partial}{\partial n} + ~\!\cos(n,r)~\!
    \bigg(\frac{\,1\,}{r} + \frac{\,1\,}{c}\frac{\partial}{\partial t}\bigg)
  \Bigg\}\,
  \Bigg(\frac{\,1\,}{s}\,\psi\big(t-s/c\big)~\!\Bigg) \,.
\end{equation}

This expression is the wave function at distance~$s$ from what I~call
a \mbox{\textbf{generalized spatiotemporal dipole}} (\textbf{GSTD}) in
the $n$ \mbox{direction}, with strength~$\psi$,\, delay factor
$\cos(n,r)$, and inverted-monopole attenuation for the distance $r$
from a monopole primary source. The \emph{operand} (on~the right,
in~the biggest parentheses) is the wave function at distance~$s$ from
a monopole of strength~$\psi\;\!$; and the composite
GSTD \emph{operator} \mbox{$\big\{$in the braces$\big\}$} can be seen
to have a spatial aspect~$\big(\frac{\partial}{\partial n}\big)$, a
temporal aspect~$\big(\frac{\partial}{\partial t}\big)$, and
``generalizations'' (delay~factor and attenuation). The same integrand
as rewritten in~\eqref{e-kand2} may then be understood as a
distribution of GSTDs, oriented normal to~$S$, the first term
representing the spatial aspect (equal and opposite monopoles) and the
second term {\large(}in~$\frac{\partial\psi}{\partial n}${\large)}
representing the ``modifications'' (delay and attenuation of the
inverted monopole).

By way of verification, we can use the chain rule
${\frac{\partial}{\partial n}\!=\!\frac{\partial s}{\partial
n}\frac{\partial}{\partial
s}\!=~\!\!\cos(n,s)~\!\frac{\partial}{\partial s}}$\, (for terms
in~$s$) to rewrite~\eqref{e-gstd} as
\begin{equation}\label{e-kdfand}
  \bigg(\frac{\cos(n,r)}{r}-\frac{\cos(n,s)}{s}\bigg)~\!\frac{[\psi]}{s}
  \,+\, \frac{\cos(n,r)-\cos(n,s)}{c}~\!\frac{\big[\dot{\psi}\big]}{s} ~,
\end{equation}
which follows similarly from~\eqref{e-kand1}, using~\eqref{e-diffs} and
${\frac{\partial}{\partial
n}\!=~\!\!\cos(n,r)~\!\frac{\partial}{\partial r}}$\, for terms
in~$r$.

According to~\eqref{e-tauh}, the delay of the inverted monopole is
such that the waves from the two monopoles are synchronized (with
opposing amplitudes) in the direction of the primary source, and in
the \emph{cone} of directions which make the same angle $(n,r)$ with
the negative direction of $n$; this cone includes the direction of
specular reflection of primary waves off $S$. And according
to~\eqref{e-alphah}, the attenuation of the inverted monopole is such
that the waves from the two monopoles cancel at a distance $r$ in any
of these directions (including at the primary source); at that
distance, the closer proximity of the inverted monopole compensates
for the reduced strength. So the GSTDs suppress backward secondary
waves in two ways: \emph{individually}, they suppress secondary waves
in particular directions, including the direction of the primary
source and the direction of specular reflection of the primary wave,
the suppression being exact at the distance of the primary
source; \emph{collectively}, they are described by the integrand
in~\eqref{e-kit} and therefore, according to the Kirchhoff integral
theorem, suppress secondary waves throughout $R'$.

Specular reflection matters because if the mathematical surface~$S$
were a partially reflective \emph{physical} surface,
the \emph{physical} secondary wavefronts emitted by each element of
the physical surface would have the same timing, relative to the
respective primary wavefronts, as the hypothetical GSTD secondary
wavefronts emitted by that element of the mathematical surface. Thus
Fermat's principle is as applicable to the mathematical surface as to
the physical one.

\section{Special cases}

Even the most general case considered above is special in that the
assumed form of the wave equation, with $c$ as a constant, implies
that the medium in~$R$ is homogeneous and isotropic.\footnote{\,\small
Combined with general time-dependence, the constancy of~$c$ also
implies that the medium in~$R$ is non-dispersive. But this restriction
can be circumvented by specializing the results for sinusoidal
time-dependence, then allowing $c$ to be frequency-dependent, and
superposing the results for all frequencies present. Accordingly, for
convenience, we press on with general time-dependence.} And the
derivation of~\eqref{e-tauh} and~\eqref{e-alphah} assumes a special
primary source, namely a single monopole. But there are two further
specializations worth mentioning.

First, if \emph{$S$\,is a primary wavefront}, as in Fresnel's
statement of Huygens' principle,\footnote{\,\small Fresnel,
tr.~Crew~\cite{fresnel-1818-crew}, at p.~\!108. Huygens himself made
no such restriction in his initial statement of the
principle \cite[p.~\!19]{huygens-1690-thompson}, although he went on
to choose secondary sources on a single primary wavefront in order to
construct the ``continuation'' of that wavefront (the same wavefront
at a later time) in the \emph{same}
medium \cite[pp.\,19,\,50--51]{huygens-1690-thompson}. To construct a
wavefront reflected or refracted at an interface between \emph{two}
media, however, he chose secondary sources at various points on the
interface, which the primary wavefront reached at various
times \cite[pp.\,23--4, 35--7, etc.]{huygens-1690-thompson}.}  then
${\cos(n,r)\!=\!1}$~\!  in~\eqref{e-tauh}, so~that $\tau_h$ becomes
$h$/$c$, which is simply the time taken for the waves emitted by the
uninverted monopole to reach the inverted monopole. The latter is in
the $-n$ direction, which is therefore the direction in which the
waves from the two monopoles are synchronized (and cancel at distance
$r$); the ``\emph{cone} of directions'' collapses to its axis.

Second, if \emph{the primary wavefronts are plane} (for a
general~$S$), we have $r\!\to\!\infty$ in~\eqref{e-alphah}, so~that
$\alpha_h\!=\!0$: the inverted monopole is not attenuated, and the
cancellation of the waves from the two monopoles [in the cone at angle
$(n,r)$ to the $-n$ direction] becomes a far-field effect.

If \emph{both} of these conditions hold\textemdash if $S$~\!coincides
with a primary wavefront \emph{and} is plane\textemdash the inverted
monopole is delayed by $h$/$c$ and is \emph{un}attenuated, so~that the
waves from the two monopoles cancel in the $-n$ direction in the far
field. The resulting dipole is what Miller~\cite{miller-91} called
a \textbf{spatiotemporal~dipole}. We~have ``generalized'' it in two
ways: by allowing the delay of the inverted monopole to be of smaller
magnitude than $h$/$c$, so~that the direction of cancellation may not
be normal to~$S$; and by allowing the inverted monopole to be
attenuated, so that the cancellation may occur at a finite
distance. Together, these modifications allow the surface of
integration $S$ to be of a general shape and orientation and at a
general distance from the primary source.

\vspace{-2ex}
\section{Approximations}
\vspace{-1ex}

Although Miller applied his spatiotemporal-dipole theory to ``uniform
spherical or plane wave fronts''~\cite[p.~\!1371, below
eq.\,5]{miller-91}, his theory is in fact a plain-wave approximation
in that it neglects the 1/$r$ decay in the magnitude of the primary
wave, with the result that his equation~(4), which corresponds to
our~\eqref{e-diffs}, lacks the second term on the right. Larmor had
done the same: his equation under the words ``and \emph{if the
surface~$S$ be a wave-front}''~\cite[p.~\!258]{larmor-1929} also lacks
that term; consequently his equation under ``and the formula
becomes''~\cite[p.~\!259]{larmor-1929}, which corresponds to
our~\eqref{e-kdfand} with integration (and notational differences),
has no term corresponding to our $\frac{\cos(n,r)}{r}$. This
approximation is valid if $S$\,is in the \textbf{far field of the
primary source}, where $r$ is much larger than any wavelength.

As $\alpha_h$ arises from the second term in~\eqref{e-diffs},\,
neglecting the attenuation of the inverted monopole amounts to
neglecting the decay of the primary wave as it
propagates.\footnote{\,\small My first attempt to ``generalize''
Miller's spatiotemporal dipole neglected~$\alpha_h\;\!$, but assumed
sinusoidal time-dependence and yielded a \emph{complex} delay, whose
imaginary part I~took as an attenuation, which I~would need to make
explicit if I~repeated the exercise with general time-dependence and
real variables. Thus the presentation of my findings does not quite
match the manner of discovery.} This is permissible not only for
spherical primary wavefronts with sufficiently large~$r$, but also
for \textbf{non-spherical primary wavefronts} whose minimum radii of
curvature are similarly large.\footnote{\,\small In a homogeneous
isotropic medium (such as the one assumed in the region~$R$),
a~non-spherical wavefront may arise from an initially plane or
spherical wavefront that has been reflected or refracted at the
interface with a different medium.} In such cases, $\cos(n,r)$ is to
be understood as the cosine of the angle between the normals of the
primary wavefront and the surface of integration.

A large-$s$ approximation, unlike a large-$r$ approximation, does not
amount to a simplification of the GSTDs, but only assumes that the
field is sufficiently \textbf{far from the GSTDs} to allow neglect of
the $\frac{\cos(n,s)}{s}$ term
in~\eqref{e-kdfand}. If~both \emph{$r$~and~$s$} are large
enough, \emph{or} the frequencies high enough, we~can neglect the
entire first term of~\eqref{e-kdfand}, leaving only the term in
$\dot{\psi}$ with the familiar Kirchhoff obliquity factor (or
``inclination factor'')\,
${\cos(n,r)\!-\!\cos(n,s)}$;\, \textit{cf}.\,\cite[p.~\!422,
eq.\,17]{born-wolf-02}. This factor is zero where the angles are
equal, including the direction of specular reflection. If, in
addition, $S$\,is a primary wavefront so that ${\cos(n,r)~\!\!=\!1}$,
while $\chi$ is the angle between the $n$ and~$-s$ directions, the
Kirchhoff obliquity factor reduces to the Fresnel--Stokes obliquity
factor
${(1\!+~\!\!\cos\chi)}$;\, \textit{cf}.\,\cite[p.~\!423]{born-wolf-02}.

Integrand~\eqref{e-gstd} is for a monopole primary source. The
integrand for a \emph{multipole} primary source\textemdash e.g., a
typical \emph{extended} source\textemdash will have a term of
form~\eqref{e-gstd} for each monopole; and for each element of the
surface~$S$, each monopole will generally give a different~$r$,
measured from a different origin.  However, if~the dimensions of the
primary source are small relative to each~$r$, then, for each surface
element, we~can take $\cos(n,r)$ and $1$/$r$, and consequently the
entire GSTD operator, as common to all terms, so~that the sum
simplifies to~\eqref{e-gstd} with $\psi$ as the total primary wave
function. Thus~\eqref{e-gstd} is approximately applicable to
a \textbf{small extended primary source}. If~$r$ is also large
compared with any wavelength, such a source will be \textbf{weakly
directional} in the sense that the variation of the primary wave
function in the tangential direction is slow compared with the
variation in the radial direction (compare single-slit interference at
a distance much larger than the wavelength and the slit width).

\section{Vector wave functions?}

The assumed form of the wave function due to a monopole secondary
source in eqs.\,\eqref{e-kand2} and~\eqref{e-uninv}, or a monopole
primary source in~\eqref{e-psimonop}, is usually taken to represent
a \emph{scalar} wave function. If~it were to represent a \emph{vector}
wave function, that vector would need to have the same direction as
the vector-valued strength function for all directions of
propagation. This requirement might seem to exclude electromagnetic
waves, for which the electric and magnetic fields are transverse to
the direction of propagation and therefore dependent on~it. However,
it is possible to describe electromagnetic waves in terms of two other
wave functions, namely an ``electric scalar potential''~$\varphi$ and
a ``magnetic vector potential''~$\mathbf{A}$, such that the
contribution to~$\mathbf{A}$ from a current element is in the same
direction as the current for all directions of propagation and has the
assumed form~\cite[pp.\,428--30]{stratton-41}. The sources of these
``potential'' waves cannot be arranged arbitrarily, because charge
must be conserved as it moves within and between current elements
(sources of~$\mathbf{A}$) and charge elements (sources
of~$\varphi$). But we need not pursue this matter further, for three
reasons. First, any realizable primary source satisfies conservation
of charge.  Second, the induced secondary sources responsible for
specular reflection are also real and therefore also satisfy
conservation of charge. Third, the secondary-source interpretation of
Kirchhoff's theorem says neither that the GSTD secondary sources
really exist, nor even that they \emph{could} exist, but only that the
wave function on the right of equation~\eqref{e-kit} is \emph{as~if}
it had been generated by such sources.

\section{Acknowledgment}

The main result of this paper was first published\textemdash or rather
buried\textemdash in a much larger work~\cite[\S~\!3.7]{putland-22},
whose content was mostly tutorial and partly historical. More recently
it was buried even deeper, in an appendix to a still-larger work on
vector analysis~\cite{putland-25}.\, I~now publish it separately in an
attempt to make it more visible and accessible.

\small\raggedright

\markright{References}


\begin{thebibliography}{9}

\bibitem{baker-copson-39} B.~\!B.~Baker \&
E.T.~Copson,~\! \textit{The Mathematical Theory of Huygens'
Principle},\, 1st~Ed., Oxford, 1939; 3rd~Ed.\ (same pagination, with
addenda), New\,York: Chelsea, 1987,
\href{https://archive.org/details/mathematicaltheo0000bake}
{archive.org/details/mathematicaltheo0000bake}.

\bibitem{born-wolf-02} M.\,Born \& E.~\!Wolf,\,
\textit{Principles of Optics}, 7th~Ed.,\, Cambridge, 1999
(reprinted with corrections, 2002).

\bibitem{fresnel-1818-crew} A.\,Fresnel,\, ``M\'{e}moire sur la
diffraction de la lumi\`{e}re'' (submitted 29~July 1818, ``crowned''
15~March 1819),\, partly translated as ``Fresnel's prize memoir on the
diffraction of light''\!,~\! in~\! H.~Crew~(ed.),~\! \textit{The Wave
Theory of Light: Memoirs by Huygens, Young and Fresnel},~\! American
Book Co., 1900, archive.org/details/wavetheoryofligh00crewrich,
\href{https://archive.org/details/wavetheoryofligh00crewrich/page/81}
{pp.\,81--144}.

\bibitem{huygens-1690-thompson} C.~Huygens (1690),\,
tr.~S.~\!P.~Thompson,\, \textit{Treatise on Light},\, Univ.\,of
Chicago Press, 1912;\, Project Gutenberg, 2005,\,
\href{https://www.gutenberg.org/files/14725/14725-h/14725-h.htm}
{gutenberg.org/files/14725/14725-h/14725-h.htm}.\, (See~also ``Errata
in various editions of Huygens' \textit{Treatise on Light}''
at \textit{www.grputland.com} or \textit{grputland.blogspot.com},
June~2016.)

\bibitem{larmor-1929} J.~Larmor,\, \textit{Mathematical and Physical
Papers}, vol.\,2,\, Cambridge, 1929;\,
\href{https://archive.org/details/mathematicalphys0002jose}
{archive.org/details/mathematicalphys0002jose}.

\bibitem{miller-91} D.~\!A.B.~Miller,\, ``Huygens's wave
propagation principle corrected''\!,\, \textit{Optics~Letters},
vol.\,16, no.~\!18 (15~Sep.\,1991), pp.\,1370--72;\,
\href{http://ee.stanford.edu/~dabm/146.pdf}
{stanford.edu/\textasciitilde{}dabm/146.pdf}.

\bibitem{putland-22} G.~\!R.~Putland,\, ``Consistent derivation of
Kirchhoff's integral theorem and diffraction formula and the
Maggi-Rubinowicz transformation using high-school math''\!,
ver.\,0.3,\, 6~Dec.\,2022.\, (Latest
version: \href{https://doi.org/10.5281/zenodo.7205781}
{doi.org/10.5281/zenodo.7205781}.)

\bibitem{putland-25} G.~\!R.~Putland,\, ``Coordinates Last: Vector
Analysis Done Fast''\!,\, \textit{Wikijournal~Preprints},
\href{https://en.wikiversity.org/wiki/WikiJournal_Preprints/Coordinates_Last:_Vector_Analysis_Done_Fast}{https://w.wiki/Ebzp}, 2025.

\bibitem{stratton-41} J.~\!A.\,Stratton,\, \textit{Electromagnetic
Theory},\, New\,York: McGraw-Hill, 1941;\,
\href{https://archive.org/details/electromagnetict0000juli}
{archive.org/details/electromagnetict0000juli}.

\end{thebibliography}
\end{document}